\documentclass[conference]{IEEEtran}
\usepackage{amsmath, amssymb,multicol, graphicx,gensymb, flushend}
\IEEEoverridecommandlockouts

\begin{document}

\title{Solid-State Diffusion: An Introduction}

\author{\IEEEauthorblockN{Jairo Rondón\IEEEauthorrefmark{1},
Angel Gonzalez-Lizardo\IEEEauthorrefmark{2},
Ricardo Bravo\IEEEauthorrefmark{3}, Juan Valera\IEEEauthorrefmark{3}}
\IEEEauthorblockA{\IEEEauthorrefmark{1}Biomedical \& Chemical Engineering Departments,\\
Polytechnic University of Puerto Rico, San Juan, Puerto Rico, USA\\jrondon@pupr.edu}
\IEEEauthorblockA{\IEEEauthorrefmark{2}Department of Electrical and Computer Engineering and Computer Science,\\
Polytechnic University of Puerto Rico, San Juan, Puerto Rico, USA\\agonzalez@pupr.edu}
\IEEEauthorblockA{\IEEEauthorrefmark{3}Biomedical Engineering Department,\\
Polytechnic University of Puerto Rico, San Juan, Puerto Rico, USA\\rbravo@pupr.edu, jvalera@pupr.edu}
%\IEEEauthorblockA{\IEEEauthorrefmark{3}Biomedical Engineering Department\\
%Polytechnic University of Puerto Rico, San Juan, Puerto Rico, USA}
}

\maketitle

\begin{abstract}
This research explores an introduction to solid-state diffusion, focusing on its importance in materials engineering. It examines vacancy and interstitial diffusion mechanisms, the application of Fick's laws, and their impact on processes such as phase precipitation and recrystallization in metals and alloys. Additionally, it addresses its relevance in grain growth, diffusion welding, and sintering, which are critical processes to improve the properties of materials with engineering applications in various areas such as biomedical, electrical, and chemistry.
\end{abstract}

\section{Introduction}
Diffusion is a fundamental phenomenon in which matter moves internally through the constant movement of atoms or molecules. This process occurs in all phases of matter: gases, liquids, and solids, although each has different characteristics and speeds. In gases, diffusion is rapid due to low density and minimal restrictions as observed in the dispersion of odors or smoke. In liquids, movement is slower due to higher density and intermolecular forces exemplified by the dispersion of ink in water. In solids, despite the tightness of atoms in a crystalline or amorphous structure, thermal vibrations allow limited but significant movement of atoms.

Solid diffusion in materials such as metals and alloys is especially important because it impacts critical processes such as secondary phase precipitation and recrystallization. These processes are essential for improving the mechanical properties of materials, from strength in aluminum alloys to ductility and toughness in cold-worked metals. Understanding the mechanisms of diffusion, whether by vacancy or interstitial, is a crucial step in optimizing these industrial processes and developing advanced materials with improved properties applicable in engineering sectors such as biomedical, electronics, and chemistry. Another significant aspect of diffusion in solids is that it considers the different states of the system, such as steady state and non-steady state, where conditions and concentrations change with time. Fick's laws provide a theoretical framework to describe these processes, allowing the prediction and control of diffusion in various industrial applications. In addition, diffusion plays a vital role in materials processing techniques such as sintering and diffusion welding, which are vital for manufacturing and improving metals, ceramics, and composites. This understanding of diffusion is fundamental to materials engineering, where precise microstructure control through heat treatments and other processes can mean the difference between a standard material and one with exceptional properties designed for specific and demanding applications.

This study explores the definition of solid-state diffusion and its mechanisms, steady-state and non-steady-state diffusion, and some elements of materials processing.

\section{Methodology}
For this research, a qualitative-documentary methodology was employed, relying on:
\begin{enumerate}
    \item Search and data compilation: Textbooks and databases such as SciELO, RedAlyC, and Google Scholar were used. Key search terms included "solid-state diffusion," "diffusion mechanisms," and "steady-state and non-steady-state diffusion."
    \item Selection and refinement of information: A comprehensive exploration was carried out using a search period from 1970 to 2023. Using Mendeley \cite{r13} as a bibliographic management tool, the data were organized based on their relevance to the four fundamental frameworks of this study: conceptualization, mechanism, diffusion status, and materials processing.
    \item Selection of subtopics: The refined information facilitated the organization of the research structure and clarified the chosen subtopics related to the study.
    \item Analysis of results: The data were critically analyzed, resulting in comprehensive conclusions.
\end{enumerate}

\section{Discussion and Results}
\subsection{Conceptualization of Diffusion in Solids}
Diffusion is a crucial phenomenon where matter moves internally through the constant movement of atoms or molecules. This process manifests itself in all phases of matter: gases, liquids, and solids. In gases, atoms and molecules move quickly due to their low density and lack of significant constraints. This rapid movement is evident in the agile dispersion of odors such as those perceived in cooking or the diffusion of smoke in the air. In liquids, particle movement is slower compared to gases. This is due to the higher density and the presence of intermolecular forces. A clear example of this phenomenon is the gradual dispersion of ink in water, where ink molecules are slowly distributed throughout the liquid. In solids, the atoms are firmly fixed in a crystalline or amorphous structure, constrained by solid bonds. Despite this, the atoms are not immobile; thermal vibrations give them the energy to move minimally from their equilibrium positions. However, this movement is much more limited than in liquids and gases.

Solid-state diffusion could then be defined as the process by which atoms, ions, or molecules move within a solid driven by concentration gradients, temperature, or other forces. This phenomenon is essential in many industrial and natural processes such as the production of materials, sintering of ceramics, creation of alloys, and hardening of metals.

Atomic diffusion in metals and alloys is particularly important because of its impact on various solid-state processes. In these materials, diffusion is essential for:
\begin{itemize}
    \item Precipitation of a second phase from a solid solution: This process is crucial to form specific microstructures that improve the mechanical properties of alloys. For example, in aluminum alloys, the precipitation of intermetallic compounds can significantly increase the strength of the material \cite{r1}.
    \item The formation of nuclei and the growth of new grains during recrystallization: In metals subjected to cold working, the internal energy of the material increases due to dislocations and other structural defects. Recrystallization, which involves the formation of new defect-free grains, decreases this internal energy and restores the metal's mechanical properties. This process is essential in producing metals and alloys with specific properties such as increased ductility and toughness.
\end{itemize}

Understanding diffusion in solids is essential to optimizing these and other industrial processes. It enables the development of advanced materials with improved properties, which have applications in a variety of sectors from chemistry to electronics and mechanics.

\subsection{Diffusion Mechanisms}
In a crystalline structure, atoms can move in two main ways: the first is through the vacancy mechanism, where atoms move to occupy the empty places left by other atoms in the lattice, and the second is through the interstitial mechanism, where smaller atoms move through the gaps between larger atoms.

\subsubsection{Vacancy or Surrogate Diffusion Mechanism}
Atoms in metals and alloys can change places within the crystal lattice if they have sufficient energy from thermal vibration. This movement is facilitated when there are vacancies or defects in the structure, which are common and in equilibrium in metals. As the temperature rises, both the vacancies and the available thermal energy increase, increasing the diffusion rate. In the case of copper, it is illustrated how its atoms move into vacancies in its crystal structure if they have sufficient energy. This necessary energy is the sum of the energy to form the vacancy and move the atom. In addition, in metals with high melting points, the binding energies between atoms are more robust, so more activation energy is required. In solid solutions, vacancy diffusion is also possible and depends on the size and binding energy differences between atoms of different elements \cite{r3}.

\subsubsection{Interstitial Diffusion Mechanisms}
Interstitial diffusion of atoms in crystal lattices occurs when atoms move from one interstice to an adjoining one without displacing any atoms in the crystal lattice matrix. For the interstitial mechanism to be effective, the size of the diffusing atoms must be relatively small compared to those in the lattice; for example, hydrogen, oxygen, nitrogen, boron, and carbon can diffuse interstitially in most metallic crystal lattices \cite{r4}.

\subsection{Diffusion in Steady State}
Consider the diffusion of solute in the x-axis direction between two atomic planes perpendicular to the plane of the paper separated by x distance. Assume that after some time the concentration of atoms in plane 1 is \(C_1\) and in plane 2 is \(C_2\). This means there is no solute concentration change with time in these planes. These diffusion conditions are known as steady state and occur when a non-reactive gas diffuses through a metal foil. For example, steady-state diffusion conditions are reached when hydrogen gas diffuses through a palladium foil if the hydrogen is at high pressure on one side and low pressure on the other.

Suppose there is no chemical interaction between the solute and solvent atoms in the system. Due to the concentration difference between planes 1 and 2, there will be a net flow of atoms from the higher to the lower concentration. The equation represents the flux or current density \cite{r10}:
\begin{equation}
J = -D \frac{dC}{dx}
\end{equation}
Where:
\begin{itemize}
    \item \(J\) is the net flux or current of atoms
    \item \(D\) is the diffusion coefficient or diffusivity coefficient
    \item \(\frac{dC}{dx}\) is the concentration gradient
\end{itemize}

The negative sign indicates that diffusion occurs from a higher concentration to a lower concentration, i.e., there is a negative gradient. This equation is called Fick's first law of diffusion and corresponds to those situations where there is no change over time \cite{r5}. The SI units for this equation are:
\begin{equation}
J \left(\frac{Atoms}{m^2 \cdot s}\right) = D \left(\frac{m^2}{s}\right) \cdot \frac{dC}{dx} \left(\frac{Atoms}{m^3} \cdot \frac{1}{m}\right)
\end{equation}

The values of the interstitial and substitutional diffusion coefficient depend on many variables, the most important of which are:
\begin{itemize}
    \item The diffusion mechanism: Small atoms can diffuse interstitially in the crystal lattice of solvents of larger atomic size, e.g., carbon in BCC or FCC phase iron. Copper atoms can diffuse substitutionally in an aluminum solvent lattice since they are approximately the same size.
    \item Temperature: The temperature at which diffusion occurs extensively affects the diffusivity value. The higher the temperature, the higher the diffusivity increases.
    \item The type of crystalline structure of the solvent: The diffusivity of carbon in \(\alpha\)-iron is \(10^{-12}\) m\(^2\)/s at 500°C, a much higher value than \(5 \times 10^{-15}\) m\(^2\)/s corresponding to carbon in \(\alpha\)-iron at the same temperature. This difference is because the BCC crystal structure has an atomic packing factor of 0.68, lower than that exhibited by the FCC structure, which is 0.74. Also, the interatomic spaces in the iron are more prominent in the BCC structure than in the FCC; therefore, carbon atoms can diffuse between the BCC iron atoms more quickly than in the FCC.
    \item Crystalline imperfections present in the region: Most open structures, e.g., grain boundaries, allow faster diffusion of atoms. Excess vacancies increase diffusion rates in metals and alloys.
    \item Concentration of diffusing elements: High concentrations of solute atoms will affect solid-state diffusion.
\end{itemize}

\subsection{Non-Steady-State Diffusion}
Steady-state, in which conditions remain invariant with time, is not a typical engineering problem. In most cases, diffusion is not stationary since the concentration of solute atoms at any point in the material changes with time. For example, suppose carbon is diffused on the surface of a steel camshaft to harden its surface. In that case, the concentration of carbon below the surface at any point will change with time as the diffusion process progresses. For cases of non-steady-state diffusion in which the diffusivity is independent of time, Fick's second law applies:
\begin{equation}
\frac{dC(x,t)}{dt} = \frac{d}{dx} \left(D \cdot \frac{dC(x)}{dx}\right)
\end{equation}

This law states that the sample composition's change rate is equal to the diffusivity times the rate of change of the concentration gradient. The derivation and solution of this differential equation are beyond the scope of this course. However, in the case of a gas diffusing in a solid, its particular solution is of great importance for some industrial processes.

Consider the case of gas A diffusing into a solid B. As time progresses, the concentration of solute atoms at any point in the solid in the x direction will increase. If the diffusivity of gas A in solid B is independent of position, then the solution to Fick's second law is:
\begin{equation}
C(x,t) = C_s \left(1 - \text{erf} \left(\frac{x}{2 \sqrt{Dt}}\right)\right) + C_0 \left(\text{erf} \left(\frac{x}{2 \sqrt{Dt}}\right)\right)
\end{equation}
Where:
\begin{itemize}
    \item \(C_s\) is the surface concentration of the element in the gas diffusing into the surface
    \item \(C_0\) is the initial uniform concentration of the element in the solid
    \item \(C(x,t)\) is the concentration of the element at distance \(x\) from the surface at time \(t\)
    \item \(D\) is the diffusion coefficient
    \item \(t\) is the time
\end{itemize}

The error function (erf) is a mathematical function used in some solutions of Fick's second law and can be found in the literature \cite{r2, r6}.

\subsection{Diffusion Processing of Materials}
Diffusion is a phenomenon of great importance in processing and improving materials, mainly when treatments are carried out at elevated temperatures. The last stages of the processing of most metals, ceramics, or polymers involve some "heat treatment" aimed at improving their mechanical, conductive, and other properties, often based on the materials' diffusion properties \cite{r8}.

\subsubsection{Grain Growth}
A material with petite grain sizes has many grain boundary regions, implying poor packing, high energy, and reactivity per unit area. One solution may be to increase the particle size to increase the stability of the material. This process implies that the grain boundaries are shifted so that some grains integrate their neighbors, which requires the atoms to diffuse across the grain boundary. Therefore, heat treatment at high temperatures, increasing the infusibility of the material, will contribute to the increase of grain sizes and decrease the material's energy \cite{r7, r11}.

\subsubsection{Diffusion Welding}
This method is used to join the same or different materials and is carried out in three stages:
\begin{itemize}
    \item Increased contact area: Materials are bonded under high pressures and temperatures to insulate surfaces and improve connectivity.
    \item Diffusion at grain boundaries: With the surfaces bonded under pressure at high temperatures, atoms diffuse down the grain boundaries into the voids at the interface, progressively reducing them. This step is rapid due to the high temperature.
    \item Volume diffusion: Slower volume diffusion will eventually fill the remaining voids.
\end{itemize}

This method is commonly used to join highly reactive metals such as titanium to other metals and ceramic components \cite{r12}.

\subsubsection{Sintering}
Sintering is crucial in powder metallurgy and manufacturing ceramic materials and composites. It allows a powder to be consolidated into a solid for subsequent manufacturing by heat treatment at temperatures below the melting point. Particles are diffusion-bonded, reducing the porosity of the material. Pressure can shorten the time required for the process. The particles contact at numerous points by compacting the powder, decreasing the porosity. Suppose the process is carried out over a long period. In that case, the material loses porosity and increases in density and grain size, which can increase the material's brittleness. To avoid this drawback, it is essential to heat under pressure to favor diffusion at lower temperatures, preventing excessive grain growth. Finally, sintering processes are also used to shorten diffusion distances and promote reactions between solids. In these cases, the solids to be reacted are ground together, pressed into a pellet to promote their contact, and heated for a specific time. The process is then regrinded and repeated several times until a complete reaction is achieved \cite{r9}.

\section{Conclusion}
Understanding and controlling diffusion processes in solids is fundamental to various industrial and scientific applications. Diffusion, defined as the movement of atoms or molecules through a medium, varies significantly between gases, liquids, and solids due to differences in density and intermolecular forces. Although atomic motion is more restricted in solids, thermal vibrations allow diffusion, which is crucial in the formation and optimization of microstructures in materials.

Diffusion in metals and alloys directly impacts processes such as secondary phase precipitation and recrystallization, which are essential to improving the mechanical properties and restoring the ductility and toughness of cold-worked materials. Diffusion mechanisms, both vacancy and interstitial, allow atomic mobility in different conditions and crystal structures, and they are influenced by temperature and the presence of crystalline defects.

In addition, Fick's laws describe steady-state and non-steady-state diffusion, providing a theoretical framework for predicting diffusion behavior in different systems. The applicability of these concepts extends to materials processing such as grain growth, diffusion bonding, and sintering, key processes for the fabrication of advanced materials with improved properties.

Finally, the study of diffusion in solids enables the optimization of industrial processes and facilitates the development of materials with innovative applications in sectors such as chemistry, electronics, and mechanics. The ability to manipulate atomic diffusion in materials remains a powerful tool in materials engineering and solids science, driving technological advances and improving the properties and functionalities of materials used in various industries.

\end{document}